\documentclass[11pt,a4paper]{article}

\usepackage{hep}
\usepackage{jheppub}
\pdfoutput=1

\date{\today}
\author{Johanna Erdmenger}
\author{and Stephan Steinfurt}
\affiliation{Max-Planck-Institut f\"ur Physik (Werner-Heisenberg-Institut),\\
F\"ohringer Ring 6, 80805 M\"unchen, Germany}
\emailAdd{jke@mppmu.mpg.de}
\emailAdd{steinfur@mppmu.mpg.de}
\title{\begin{center}A universal fermionic analogue of the shear viscosity\end{center}}
\abstract{
We holographically compute supercharge diffusion constants in supersymmetric field theories, dual to AdS black brane solutions of arbitrary dimension. This includes the extension of earlier work by Kontoudi and Policastro for D3-branes to M2- and M5-brane theories. We consider the case of vanishing chemical potential. In particular, we relate the product of a diffusion constant and the energy density to a universal result for the fermionic absorption cross section. This relation is analogous to the famous proof of universality of $\eta / s$. We calculate the diffusion constants in two different ways: First, the computation is performed via the low frequency, low momentum pole of the correlator of supersymmetry currents. This pole describes the hydrodynamic phonino mode, the massless Goldstone fermion of spontaneous supersymmetry breaking by temperature. Second, the calculation is carried out using the dual transversal mode of the bulk gravitino, with the help of a new Kubo formula. Moreover, we provide some 
evidence for the applicability of generalized dimensional reduction for fermions when computing the corresponding D$p$-brane diffusion constants.
}

\preprint{MPP-2013-10}

\begin{document}
\maketitle

\section{Introduction}
Holographic hydrodynamics has been very successful at calculating transport coefficients for strongly coupled systems, in particular with regard to applications to real-world systems such as the quark-gluon plasma~\cite{Policastro:2002se,Herzog:2002fn,Son:2007vk,Hubeny:2010ry,CasalderreySolana:2011us}. One of the most far-reaching results obtained in this context is the universality of the ratio of shear viscosity and entropy density $\eta / s = 1/ 4 \pi$ for theories which have a dual description in terms of Einstein gravity with unbroken rotational symmetry~\cite{Buchel:2003tz,Kovtun:2004de,Buchel:2004qq,Benincasa:2006fu,Son:2007vk,Iqbal:2008by}.

A further recent line of investigation within gauge/gravity duality concerns the study of fermions. In particular, new results for fermionic correlators have been found in models which describe strongly coupled systems that are interesting in view of applications to condensed matter physics~\cite{Lee:2008xf,Liu:2009dm,Cubrovic:2009ye,Faulkner:2009wj}~\cite{McGreevy:2009xe,Hartnoll:2009sz,Iqbal:2011ae}. Such systems have also been studied from a top-down perspective in~\cite{Ammon:2010pg,Jensen:2011su} or more recently in~\cite{DeWolfe:2011aa,DeWolfe:2012uv}. This naturally leads to the question if from these correlators we may find universal results which are similar to the universality of $\eta / s$.

A candidate which may be computed from fermionic correlators has been studied and speculated to have universal properties in~\cite{Policastro:2008cx}\footnote{For other approaches towards a holographic description of a seemingly universal behaviour of holographic superconductors see~\cite{Erdmenger:2012ik}.}. On general grounds, in the low energy, low momentum limit any quantum field theory at finite temperature may effectively be described by hydrodynamics~\cite{Landau:1959}. In the hydrodynamical limit, the shear viscosity may then be calculated from the two-point function of the theory's transverse energy momentum tensor $\left<T_{xy}T_{xy}\right>$ by applying a Kubo formula. In supersymmetric field theories there is also the supersymmetry current $S_\mu^\alpha$, which belongs to the same supermultiplet as the energy-momentum tensor. Therefore we may wonder if in the hydrodynamic limit of a supersymmetric field theory the two-point function of the supersymmetry current also possesses a similar 
universality. However, since supersymmetry is broken spontaneously by temperature the relation between $\left<T_{\mu\nu}T_{\rho\sigma}\right>$ and $\left<S_\mu \bar{S}_\nu\right>$ is non-trivial. For theories at finite temperature the expectation value $\left<T_{\mu\nu}\right>$ does not vanish, so from inspection of the Ward identity $\partial^\mu \left< TS_\mu(x) \bar{S}_\nu(0)\right> = \delta^{(4)}(x)\,2\,\gamma^\mu \left<T_{\mu\nu}\right>$~\cite{deWit:1975th,deWit:1975nq}, we see that there must be a pole in $\left<S_\mu\bar{S}_\nu\right>$ describing the Goldstone fermion of spontaneously broken supersymmetry~\cite{Leigh:1995jw}. This pole may be interpreted as the so-called phonino mode~\cite{Lebedev:1989rz,Kratzert:2003cr,Kratzert:2002gh}, a massless excitation with characteristic sound dispersion relation $\omega = v_s k - i D_s k^2$ whose appearance resembles that of the phonon. This was first established holographically for $\mathcal{N}=4$ SYM in 4 dimensions in~\cite{Policastro:2008cx}, where the 
concrete dispersion relation was computed analytically. In~\cite{Policastro:2008cx}, the equations of motion of a gravitino in AdS$_5$ were solved to linear order in $\omega$ and $k$ to derive the retarded Greens functions of the dual supersymmetry current operators using the prescription of~\cite{Son:2002sd,Policastro:2002se}, from which the dispersion relation may be read off. This setup has been studied further in~\cite{Kontoudi:2012mu} using the transverse mode of the gravitino and a Kubo formula~\cite{Kovtun:2003vj}, also for non-vanishing chemical potentials. In three space-time dimensions, the correlator of supersymmetry currents at finite chemical potential has been studied in~\cite{Gauntlett:2011mf,Belliard:2011qq,Gauntlett:2011wm} and, interestingly, it was shown that no Fermi surfaces show up in the spectral functions of their setups.

The quantity which has been speculated to have similar universal properties as the shear viscosity $\eta$ is the supersound diffusion constant $D_s$, which also appears as a particular transport coefficient in the constitutive relation of the supersymmetry current. In four dimensions, this is given by~\cite{Kovtun:2003vj}
\begin{equation}
    S^i_{\textrm{diss}} = - D_s \nabla^i \rho - D_\sigma  \sigma^{ij} \nabla_j \rho\label{dissipative_constitutive_relation_supercurrent}
\end{equation}
with $\rho$ the supercharge density. 

However, as emphasized in~\cite{Hoyos:2012dh}, this supersymmetric hydrodynamics should really be understood as the low-energy effective theory of the phonino moving in the normal fluid. 
Since expectation values for fermionic operators vanish, one should not introduce a classical (fermionic) supercharge density in terms of which constitutive relations for the spatial parts of the supersymmetry current are expanded, as is done for normal hydrodynamics with conserved (bosonic) charges. Instead, the supercharge density has to be interpreted as the quantum phonino field itself. In addition, according to~\cite{Hoyos:2012dh}, the fermionic chemical potentials have to be viewed as external gravitino sources, where $D_s$ and $D_\sigma$ are interpreted as masses for the spin 3/2 and spin 1/2 components of these. This approach does not alter the form of the constitutive relations as compared to~\cite{Kovtun:2003vj}, but their interpretation is refined in the sense that the first derivative terms should not be thought of as dissipative parts contributing to the entropy current.

Furthermore, there is a second motivation for studying the two-point function of the supersymmetry current. Originally, the AdS / CFT correspondence~\cite{Maldacena:1997re,Gubser:1998bc,Witten:1998qj} has arisen from the study and comparison of classical absorption cross sections in black brane backgrounds and decay rates of external modes decaying into the worldvolume fields on D$p$-branes~\cite{Klebanov:1997kc,Gubser:1997yh,Gubser:1997se}. When studying such classical absorption cross sections in detail, it had before been realized that the low-energy s-wave absorption cross section of minimally coupled massless scalars is universally given by the area of the horizon of the considered black brane background~\cite{Das:1996we}. This very result was then used later to show the universality of $\eta / s$ in the holographic context~\cite{Kovtun:2004de}. This proof relies on the fact that the metric fluctuation for the shear mode $h_{xy}$ generically satisfies $\square h_x^y=0$. Moreover, the Kubo formula and 
absorption cross section formula are basically identical in the low-energy limit~\cite{Policastro:2001yc}. In~\cite{Das:1996we}, a further universal absorption cross section result has been given for minimally coupled massless fermions. We may therefore hope that the transverse gravitino, which is the dual mode to the supersymmetry current, also satisfies the equation of motion of a spin 1/2 fermion. Then the universality of the fermionic absorption cross section would lead to a universal relation which involves the supersound diffusion constant in very much the same way as the universality of $\eta / s$ can be seen as originating from the universal absorption cross section result for scalars.

In this paper, we therefore study the supersound diffusion constant $D_s$ by computing it in various AdS / CFT setups of arbitrary dimension which include D3-, M2- and M5-brane theories. We then search for and actually show universality in the way just described. Our main result therefore does indeed relate a hydrodynamic transport coefficient to a universal absorption cross section. This universal relation is given by
\begin{equation}
   {\epsilon} D_{3/2} = \frac{1}{4\pi G}\,\sigma_\text{abs,1/2}(0)\,.\label{universal_relation}
\end{equation}
On the left hand side, there is the product of the energy density $\epsilon$ of the field theory and a transport coefficient which we refer to as $D_{3/2}$. The latter appears as a specific transport coefficient in the dissipative part of the constitutive relation for the supersymmetry current after reordering the $d$-dimension generalization of~\eqref{dissipative_constitutive_relation_supercurrent},
\begin{equation}
    S^i_{\textrm{diss}} = - D_{3/2} \left(\delta^i_j-\frac{1}{d-1} \gamma^i \gamma_j\right)\nabla^j \rho - D_{1/2} \gamma^i \slashed{\nabla} \rho\,.
\end{equation}
On the right hand side of~\eqref{universal_relation}, there is the low energy absorption cross section of a minimally coupled massless spin 1/2 fermion by the dual gravity background, where the transverse space, e.g.\ for AdS$_5\times$S$^5$ the five-sphere, has not been reduced on. This relation is analogous to the corresponding one for the shear viscosity,
\begin{equation}
 \eta = \frac{1}{16 \pi G}\,\sigma_\text{abs,0}(0)\,.
\end{equation}
In the latter case we have $\sigma_\text{abs,0}(0)=a$, the area (density) of the horizon, and therefore obtain the universality of $\eta / s$ from the Bekenstein-Hawking entropy relation. In the fermionic case the absorption cross section is also related to the area of the horizon of a gravity background, however not of the dual background, but of a different, conformally related background to be described in detail in section~\ref{motivation}. Therefore, we cannot as straightforwardly divide by the entropy density of the field theory to obtain a universal relation which is as powerful as $\eta / s = 1 / 4\pi$. However, the relation~\eqref{universal_relation} is still striking. It appears possible that a universal result similar to $\eta / s = 1 / 4\pi$ may be obtained by dividing not by the entropy density itself, but by a closely related quantity.

This calculation for asymptotically AdS black brane spacetimes might also be used to compute the corresponding supersound diffusion constants for non-conformal D$p$-brane theories. In~\cite{Kanitscheider:2009as,Gouteraux:2011qh} the technique of `generalized dimensional reduction' has been developed, which allows for computing e.g.\ shear and bulk viscosity of specific non-conformal theories from using results of related conformal ones. The relation between D$p$-brane and asymptotically AdS black brane theories is exactly of such a generalized dimensional reduction type. Therefore, it seems natural to use it, although we have to point out that the details of this relation in the case of fermions have not been worked out so far.

The paper is organized as follows. In section~\ref{cross_sections} we start by giving more details on the relation between hydrodynamic transport coefficients in the field theory and universal absorption cross sections in the corresponding gravity dual by first reviewing the arguments of~\cite{Kovtun:2004de,Das:1996we} and then extending them to the fermionic case. We moreover derive new Kubo formulae for $D_{3/2}$, which are very similar to the ones given in~\cite{Kovtun:2003vj,Kontoudi:2012mu}, but may be directly related to the classical absorption cross section results for spin $1/2$ particles. We then apply the results to a family of asymptotically AdS black brane space-times~\cite{Gibbons:1994vm,Gibbons:1993sv} which includes D3-, M2- and M5-branes and derive their supersound diffusion constant in two mutually consistent ways. In section~\ref{transverse} we apply the methods of~\cite{Kontoudi:2012mu} to these backgrounds using the earlier Kubo formulae. We then go on and use the technically more 
involved methods of~\cite{Policastro:2008cx} in section~\ref{longitudinal} and also arrive at the same result for $D_s$, but furthermore find the expected supersound velocities $v_s$. The final section~\ref{Dp_branes}, before we conclude in section~\ref{conclusion}, uses the given results to derive the corresponding values of the supersound diffusion constant for the non-conformal D$p$-brane backgrounds. Here, the technique of generalized dimensional reduction~\cite{Kanitscheider:2008kd,Kanitscheider:2009as} is applied. Although not studied for fermions yet, some evidence in favour of the applicability of this approach is given.

\section{Relation to black hole absorption cross sections}\label{cross_sections}

In this section we derive a new universal result for a hydrodynamic transport coefficient similar to the universality of the ratio of shear viscosity $\eta$ over entropy density $s$. This result will be used to rederive the supersound diffusion constant $D_s$ (at vanishing chemical potential for R charges) for the D3-brane theory~\cite{Policastro:2008cx,Kontoudi:2012mu}. Moreover we extend this approach to also include the M2- and M5-brane theories. Furthermore, our approach also applies to a whole class of near-extremal non-dilatonic $p=d-1$ branes in AdS$_{d+1}$~\cite{Horowitz:1998ha} which arise from the near-horizon limit of a class of $D$-dimensional black $p$-brane solutions~\cite{Gibbons:1994vm,Gibbons:1993sv} and a subsequent sphere reduction.

\subsection{Motivation}\label{motivation}
One of the prime results in holographic hydrodynamics have been the universality proofs~\cite{Buchel:2003tz,Kovtun:2004de,Buchel:2004qq,Benincasa:2006fu,Son:2007vk,Iqbal:2008by} of the quantity
\begin{equation}
   \frac{\eta}{s}=\frac{1}{4\pi}\label{KSS_result}\,.
\end{equation}

In one of these proofs~\cite{Kovtun:2004de}, which we will quickly review here, it is assumed that the thermal field theory's holographic dual description is in terms of fluctuations about a black brane background with transverse $O(2)$ symmetry. Then the shear viscosity $\eta$ of the hydrodynamic limit of the field theory with energy momentum tensor $T_{\mu\nu}$ is related to the low-energy absorption cross section $\sigma_\text{abs,0}(\omega=0)$ of a transverse bulk graviton $h_{xy}$ by this brane. Comparing the Kubo formula for $\eta$ 
\begin{equation}
   \eta = \lim_{\omega\rightarrow 0} \frac{1}{2\omega}\int d^d x\, e^{i\omega t} \left\langle \left[T_{xy}(x),T_{xy}(0)\right]\right\rangle\label{Kubo_eta}
\end{equation}
with the holographic result for the low-energy absorption cross section~\cite{Klebanov:1997kc,Gubser:1997yh}
\begin{equation}
   \sigma_\text{abs,0}(\omega)=-\frac{2\kappa^2}{\omega} \textrm{ Im } G^R(\omega)=\frac{\kappa^2}{\omega}\int d^d x \,e^{i\omega t} \left\langle \left[T_{xy}(x),T_{xy}(0)\right]\right\rangle\label{sigma_correlator_scalar}
\end{equation}
it was shown~\cite{Policastro:2001yc} that
\begin{equation}
 \eta = \frac{1}{16 \pi G}\,\sigma_\text{abs,0}(0)\,.\label{PSS_observation}
\end{equation}

Since the entropy of the black brane is given in terms of the area of its horizon $S=A/4G$ ($s$ and $a$ are the corresponding densities), we may then use that the low-energy absorption cross section of $h_x^y$, which obeys the equation of motion of a massless minimally coupled scalar $\square h_x^y=0$, is given by the area of the horizon
\begin{equation}
  \sigma_\text{abs,0}(0)=a\label{sigma_scalar}
\end{equation}
to show the result of~\eqref{KSS_result}. 

The cross-section result~\eqref{sigma_scalar} had previously been proved by Das, Gibbons and Mathur~\cite{Das:1996we} for arbitrary dimensional spherically symmetric black hole backgrounds,
\begin{equation}
 ds^2=-f(r)dt^2 + g(r)\left(dr^2 + r^2 d\Omega_p^2\right)\, .\label{DGM_black_hole}
\end{equation}
For the proof, these backgrounds are assumed to have a horizon at $r=R$, to be asymptotically flat and non-extremal. This is a slightly unusual coordinate system compared to more standard ways to write the background e.g.\ of a black brane in AdS$_5\times$S$^5$, in which there is no warp factor such as $g(r)$ in front of the \emph{transverse} sphere. Nevertheless, this is just a coordinate change. The proof for $\sigma_\text{abs,0}(0)=a$ then involves solving the equation of motion of a massless minimally coupled scalar in the regions far away and close to the horizon, matching these exact asymptotic solutions and from this calculate the absorption probability of the s-wave and the absorption cross section as in~\cite{Unruh:1976fm}. This also holds for charged and/or rotating black holes~\cite{Maldacena:1997ih} and can be extended to branes~\cite{Emparan:1997iv}.

Furthermore, in~\cite{Das:1996we} a similar result for the low-energy absorption cross section of a massless minimally coupled fermion by the black hole~\eqref{DGM_black_hole} has been obtained,
\begin{equation}
  \sigma_\text{abs,1/2}(0)=2\, g_H^{-p/2}\,a\,,\label{sigma_fermion}
\end{equation}
where $g_H$ is $g(r)$ of~\eqref{DGM_black_hole} evaluated at the horizon $R$ and the factor of $2$ comes from the two helicities of the spinor. Although this is a coordinate dependent result, it is universal in the sense that it is twice the area of the horizon evaluated in the conformally related spatially flat metric $ds^2=dr^2 + r^2 d\Omega_p^2$. Furthermore, note that this low-energy absorption cross section is determined by \emph{horizon quantities} only!

We may use~\eqref{sigma_fermion} to efficiently calculate the absorption cross sections of minimally coupled massless fermions for diverse backgrounds~\cite{Jung:2005sw}. For example, for the four-dimensional asymptotically flat Schwarzschild geometry we may easily show $\sigma_\text{abs,1/2}=1/8 \,\sigma_\text{abs,0}$ in agreement with Unruh's classic result~\cite{Unruh:1976fm}.

For vanishing R-charge chemical potential, the transverse gravitino satisfies the equation of motion of a minimally coupled spin 1/2 fermion, as we will later show in a particular example~\eqref{eom_eta} and is a direct consequence of gauge invariance. So, we may wonder if this simplification along with~\eqref{sigma_fermion} also implies a universality relation as in~\cite{Kovtun:2004de}.

What is then the relation analogous to~\eqref{PSS_observation} and what would be the universal quantity as in~\eqref{KSS_result}?

\subsection{Fermion absorption cross section}
To obtain analogous fermionic results, let us first derive a relation for the absorption cross section $\sigma_\text{abs,1/2}(\omega)$ of a bulk fermion similar to the scalar result~\eqref{sigma_correlator_scalar}, which will turn out to be very useful. The slight complication, however, in generalizing~\eqref{sigma_correlator_scalar} to particles with spin is that we have to specify a polarization of the infalling particle. 

From the field theory point of view, this bulk absorption may be seen as the decay of a massive particle into the world-volume theory's fields it couples to. Then the absorption cross section is just given by the standard formula of the field theory decay rate,
\begin{align}
   \sigma_{1/2}=\frac{1}{2\omega}\int d\Pi\, |\mathcal{M}|^2\,,
\end{align}
where $\int d\Pi$ denotes the final state particles' momentum space integrals including the overall momentum conserving delta function.
Let us take the bulk Dirac fermion $\Psi$ to have the following kinetic term\footnote{The non-canonical normalization will be explained later, following eq.~\eqref{bdy}.}, 
\begin{equation}
    \frac{4}{\kappa_{d+1}^2}\int  d^{d+1}x \sqrt{-g}\,\bar{\Psi}\slashed{D}\Psi \label{non_canonical_bulk_fermion}
\end{equation}
and couple its boundary values $\Psi_0$ and $\bar{\Psi}_0$ to a spinorial boundary operator $S$, which will later turn out to be a specific transverse component of the supersymmetry current, via
\begin{equation}
\int d^dx \left(\bar{S}P_-\Psi_0 + \bar{\Psi}_0P_+S\right)\,.
\end{equation}
Here $P_\pm =\frac{1}{2}\left(1\pm\gamma^{d}\right)$, in which $\gamma^d$ denotes the gamma matrix corresponding to the radial AdS coordinate. Then for even $d$, $P_-\Psi_0$, $\bar{\Psi}_0P_+$, $\bar{S}P_-$ and $P_+S$ are chiral spinors as seen from the boundary theory, while for odd $d$, they are all Dirac boundary fermions~\cite{Henningson:1998cd} (see also~\cite{Iqbal:2009fd}).

Then, we may use the optical theorem $2 \,\text{Im} \mathcal{M} = \int d\Pi\, |\mathcal{M}|^2$ and average over the polarizations for the decaying Weyl / Dirac spinors in even / odd $d$. Furthermore, one can use a corresponding spin sum identity $\sum u\bar{u}=\slashed{p}+m$. Since our decaying fermion is at rest $p^\mu=(\omega,\vec{p}=0)$ from the point of view of the boundary theory, only the $\gamma^0$ part is left over,
\begin{align}
   \sigma_\text{abs,1/2}(\omega)&=\frac{\kappa_{d+1}^2}{\text{ Tr}\left(-\gamma^0 \gamma^0\right)}\text{ Tr}\left(-\gamma^0  \text{ Im}\int d^dx \,e^{i \omega t} \,\left\langle \,P_+ S(x) \bar{S}(0) P_- \,\right\rangle\right) \text{ for even } d \label{abs_fermion_1}\\
   \sigma_\text{abs,1/2}(\omega)&=\frac{\kappa_{d+1}^2}{2\text{ Tr}\left(-\gamma^0 \gamma^0\right)}\text{ Tr}\left(-\gamma^0 \text{ Im} \int d^dx\, e^{i \omega t} \,\left\langle \,S(x) \bar{S}(0)\,\right\rangle \right) \text{ for odd } d \, .\label{abs_fermion_2}
\end{align}

The given absorption cross section formulae are closely related to the Kubo formulae for a specific hydrodynamic transport coefficient, as we are about to explain now.

\subsection{Constitutive relation and Kubo formula}

A generalization of the four-dimensional constitutive relation which relates the spatial part of the supersymmetry current $S^i$ to the supercharge density $\rho = S^0$ is\footnote{Note that, as pointed out in~\cite{Hoyos:2012dh}, we should really understand $\rho$ as the phonino quantum field which then determines the supercharge density.}
\begin{equation}
   S^i = \frac{P}{\epsilon} \gamma^i \gamma^0 \rho - D_s \nabla^i \rho +  \frac{D_\sigma}{(d-2)} \gamma^{[i}\gamma^{j]} \nabla_j \rho\,,\label{constitutive_relation}
\end{equation}
where $\epsilon$ and $P$ are the energy density and pressure of the fluid. The non-dissipative part is fully determined by the $d$-dimensional supersymmetry algebra~\cite{Strathdee:1986jr}. $D_s$ and $D_\sigma$ are transport coefficients determining the damping of a sound-like excitation, the phonino, which propagates at the speed $v_{s}=\frac{P}{\epsilon}$~\cite{Lebedev:1989rz,Leigh:1995jw}. In the superconformal case we have $\gamma_\mu S^\mu =0$, since supersymmetry relates this to $T^\mu_\mu=0$, and therefore $D_s=D_\sigma$ and $v_s=\frac{1}{d-1}$.

For minimal supersymmetry in $d=4$ dimensions, we may recover the Weyl form of this constitutive relation~\cite{Kovtun:2003vj} by taking $\rho$ to be a Majorana spinor in the Weyl basis. Depending on the dimension, we however take $\rho$ to be a Dirac spinor (when $d$ is odd), or we project to the Weyl version of~\eqref{constitutive_relation} (for $d$ even) with $\rho$ and $\bar{\rho}$ to denote Weyl spinors. As already mentioned, this is convenient since these types of spinors are the boundary spinors inherited from Dirac spinors in $d+1$ bulk dimensions, which are the easiest to handle in general dimension. Of course this means that, depending on the dimension, the supercharge will not give rise to minimal but rather extended supersymmetry (see e.g.~\cite{Gauntlett:2011mf}).

We may reorder the constitutive relation~\eqref{constitutive_relation} according to the spinorial representations of $O(d-1)$,
\begin{equation}
   S^i =\frac{P}{\epsilon} \gamma^i \gamma^0 \rho - D_{3/2} \left(\delta^i_j-\frac{1}{d-1} \gamma^i \gamma_j\right)\nabla^j \rho - D_{1/2} \gamma^i \slashed{\nabla} \rho\,,
\end{equation}
where $D_{3/2} = D_s + \frac{1}{d-2} D_\sigma$ and $D_{1/2} = \frac{1}{d-1} \left(D_s -  D_\sigma\right)$ are the transport coefficients corresponding to the spin $3/2$ and spin $1/2$ parts under $O(d-1)$ of the vector spinor $\nabla^j \rho$. This way to write the constitutive relation is completely analogous to the way the energy momentum tensor is conventionally written involving the shear and bulk viscosities $\eta$ and $\zeta$. These transport coefficients appear in front of the symmetric traceless and trace parts of $\nabla^i u^j$ (where $u^j$ is the fluid velocity) in the first order dissipative part of the energy momentum tensor. In the conformal case, we have
\begin{equation}
   D_s = \left(\frac{d-2}{d-1}\right) D_{3/2} \quad\text{and}\quad D_{1/2} =0\,.\label{conformal_diffusion_constants}
\end{equation}
Basically these redefinitions are entirely equivalent to writing the usual sound attenuation in terms of shear and bulk viscosity \cite{Policastro:2002tn,Herzog:2003ke}. Now $D_{3/2}$ may be calculated via a Kubo formula. For even $d$ we have
\begin{align}
   \epsilon D_{3/2}&=\frac{2}{\text{ Tr}\left(-\gamma^0 \gamma^0\right)}\left(\frac{1}{d-2}\right)\lim_{\omega,k\rightarrow 0}\text{ Tr}\left(-\gamma^0 \text{ Im} \int d^dx \,e^{i \omega t} \,\left\langle \,P_+ S^i_T(x) \bar{S}^i_T(0)P_- \,\right\rangle\right)\,. \label{Kubo}
   \intertext{For odd $d$ we get}
   \epsilon D_{3/2}&=\frac{1}{\text{ Tr}\left(-\gamma^0 \gamma^0\right)}\left(\frac{1}{d-2}\right)\lim_{\omega,k\rightarrow 0}\text{ Tr}\left(-\gamma^0 \text{ Im} \int d^dx\, e^{i \omega t} \,\left\langle \,S^i_T(x) \bar{S}^i_T(0)\,\right\rangle \right) \,,\label{Kubo2}
\end{align}
where the limit $k\rightarrow 0$ is taken first as usual. The derivation is very similar to the one outlined in the appendix A of~\cite{Kontoudi:2012mu}: We use the solution to the current conservation equation $\partial_\mu S^\mu=0$ involving the constitutive relation~\eqref{constitutive_relation} to obtain the correlator $\left\langle \rho \bar{\rho}\right\rangle$. This then determines the correlator $\left\langle S^i_T \bar{\rho}\right\rangle$, where 
\begin{equation}
    S^i_T=\left(\delta^i_j - \frac{1}{d-1}\gamma^i\gamma_j\right)S^j=-D_{3/2}\left(\delta^i_j - \frac{1}{d-1}\gamma^i\gamma_j\right)\nabla^j\rho
\end{equation}
denotes the spin $3/2$ part of the supersymmetry current under the spatial $O(d-1)$. From this we get
\begin{equation}
\text{Im}\,\left(\frac{k_i}{k^2}\,\left\langle S^i_T \bar{\rho}\right\rangle\right)=-\left(\frac{d-2}{d-1}\right)D_{3/2}\,\text{Re}\, \left\langle \rho \bar{\rho}\right\rangle\,.
\end{equation}

Using Ward identities and appropriate limits, we may then turn the current-charge correlator into the current-current correlators~\eqref{Kubo} and~\eqref{Kubo2}. Note that the normalizations in~\eqref{Kubo} and~\eqref{Kubo2} do agree with the expectation since at vanishing $\vec{k}$, from rotational invariance, we have $\left\langle S^i \bar{S}^j\right\rangle \propto \delta^{ij}$. Therefore, $\delta^{j}_i\left(\delta^i_j - \frac{1}{d-1}\gamma^i\gamma_j\right)=d-2$ for $i,j=1,\ldots,d-1$ gives the correct number of independent modes of a vector $S^i_T$ keeping in mind the imposed constraint $\gamma_i S^i_T=0$ from the projection. Similar reasoning regarding the purely spinorial degrees of freedom may also be applied. For these, the normalizations of the Kubo formulae are expected to involve expressions like $\text{Tr}\left(-\gamma^0 \gamma^0\right)=2^{\lfloor\frac{d}{2}\rfloor}$ in the Dirac or $\frac{1}{2}\text{Tr}\left(-\gamma^0 \gamma^0\right)$ in the Weyl case.

At non-vanishing $\vec{k}$, one would use the projector $P_{ij}^T$, which projects onto the $\gamma^i$-traceless, $\vec{k}$-transverse part of a vector-spinor
\begin{equation}
   P^T_{ij}=\delta_{ij}-\frac{1}{d-2}\left(\gamma_i-\frac{k_i \slashed{k}}{k^2}\right)\gamma_j - \frac{1}{(d-2)k^2}\left((d-1) k_i - \gamma_i\slashed{k}\right)k_j\,,\label{transverse_projector}
\end{equation}
again very similar to the shear viscosity case.

Since the gravitino couples to the supersymmetry current on the boundary, using gauge/gravity duality we may relate the gravitino absorption cross section by the brane to the retarded Green's function of the dual operator. This is then a fermionic analogue of the graviton absorption cross section considered in~\cite{Gubser:1997yh}. Similar to the way one considers transverse metric perturbations $h_{xy}$ for the $\eta / s$ case, we here focus on the gravitino modes which have spin $3/2$ under the $\vec{k}$ preserving little group $O(d-2)$. These are referred to as $\eta_i=P^T_{ij}\Psi^j$. For vanishing R-charge chemical potentials these transverse gravitino components satisfy equations of motion of minimally coupled fermions similar to the transverse graviton obeying a Klein-Gordon equation. We may therefore relate the absorption cross section results~\eqref{abs_fermion_1}~\eqref{abs_fermion_2} to the Kubo formulae~\eqref{Kubo} and~\eqref{Kubo2}, in which $O(d-1)$ symmetry at vanishing $\vec{k}$ implies 
$\frac{1}{d-2} S^i_T \bar{S}^i_T = S^x_T \bar{S}^x_T\equiv S \bar{S}$. Putting these together, we obtain
\begin{equation}
 {\epsilon} D_{3/2} = \frac{2}{\kappa_{d+1}^2}\,\sigma_\text{abs,1/2}(0)= \frac{1}{4\pi G}\,\sigma_\text{abs,1/2}(0)\,,\label{Kubo_sigma}
\end{equation}
which is the fermionic analogue of~\eqref{PSS_observation}.

However the bulk fields $\eta_i$ are not massless as required for the use of~\eqref{sigma_fermion}, but rather have mass, for instance $m l = \frac{d-1}{2}$ in AdS$_{d+1}$. Therefore we cannot directly relate~\eqref{Kubo_sigma} to the universal result~\eqref{sigma_fermion}, but rather need to generalize~\eqref{sigma_fermion} to massive fermions.

On the other hand, the AdS fermions $\eta_i$ get their mass $m\sim \frac{1}{l}$ only from (consistent) Kaluza-Klein reduction on the transverse sphere, albeit being massless from the point of view of the higher-dimensional theory. We may therefore still compute the absorption cross section in the higher-dimensional theory, directly using the $m=0$ results~\eqref{sigma_fermion}. We will pursue both paths in the following subsections.

So far, we do not quite understand the field theory meaning of the right hand sides of~\eqref{sigma_fermion} and therefore~\eqref{Kubo_sigma} in a holographic context. Since the absorption cross section is closely related to the horizon area and therefore entropy density $s$, we would like to divide by a quantity like $s$ on both sides. However the right hand side of~\eqref{sigma_fermion} is the area of the horizon in a conformally related spatially flat metric and not the horizon area measured in the original metric. It would be very interesting to understand this better.

\subsection{Massive / higher dimensional absorption cross section}
For computing the effect of mass on the absorption cross section result~\eqref{sigma_fermion}, we very closely follow~\cite{Das:1996we} in their derivation of~\eqref{sigma_fermion}, but start off with the \emph{massive} Dirac equation for a minimally coupled spinor field $\Psi$,
\begin{equation}
\left(\nabla_\mu \gamma^\mu - m\right)\Psi =0\,.
\end{equation}

It is easy using the conformal properties of the Dirac equation (under $g_{\mu\nu}\rightarrow \Omega^{-2}g_{\mu\nu}$ in $d$ dimensions one has $(\Psi,m)\rightarrow (\Omega^{\frac{d-1}{2}}\Psi, \Omega m)$, cf.~\cite{Gibbons:1993hg}) to show that in the background~\eqref{DGM_black_hole} this is equivalent to
\begin{equation}
 h \gamma^i\partial_i  \chi =i\omega \gamma^0 \chi+m f^{1/2} \chi
\end{equation}
for the spinor $\chi =  f^{1/4} g^{p/4}\Psi$ and $h=\sqrt{f/g}$ and $\gamma^i\partial_i=\gamma^r[\partial_r+\frac{p}{2r}]+\frac{1}{r}(\gamma^i\nabla_i)_T$. Using a basis for spinors that satisfies $\gamma^{r}\lambda^\pm_n=\pm \lambda^\pm_n$ and $\gamma^{0}\lambda^\pm_n=\mp \lambda^\mp_n$, we may expand
\begin{equation}
  \chi =\sum_{n=0}^\infty F_n(r) \lambda_n^+ + G_n(r)\lambda_n^-
\end{equation}
and use the known spectrum of the Dirac operator on the sphere $\gamma^r(\gamma^i\nabla_i)_T \lambda_n^\pm = \mp (n+\frac{p}{2})\lambda_n^\pm$ (see e.g.~\cite{Camporesi:1995fb}) to arrive at
\begin{subequations}
  \label{massive_Dirac_equation_FG}
  \begin{align}
  h\left(\partial_r - \frac{n}{r}\right)F_n - m f^{1/2} F_n &= i \omega G_n\,,\\
  h\left(\partial_r + \frac{p+n}{r}\right)G_n + m f^{1/2} G_n &= i \omega F_n\,,
  \end{align}
\end{subequations}
which only slightly modifies the equations of~\cite{Das:1996we}. Now, eliminating $G_n$ one will for the $n=0$ mode of the spinor end up with
\begin{equation}
 h\left(\partial_r+p/r+m\sqrt{g}\right)h\left(\partial_r - m\sqrt{g}\right)F_0 + \omega^2 F_0 =0\,.
\end{equation}
We may now define a new coordinate via $\frac{d}{dx} = h(r)r^p \rho(r) \frac{d}{dr}$ under the condition $\rho^{-1}\partial_r \rho=2m\sqrt{g}$ and $\rho\rightarrow 1$ as $r\rightarrow \infty$. Then defining $F_0 = \exp\left(m\int dr\sqrt{g} \right)\tilde{F}$ we have
\begin{equation}
 \partial_x^2\tilde{F}+\omega^2 r^{2p}\rho^2 \tilde{F}=0\,,
\end{equation}
which is of a suitable form to compare with the scalar case. Following the arguments in~\cite{Das:1996we}, so choosing an ingoing wave at the horizon, we directly obtain that the absorption cross section for a minimally coupled massive spin 1/2 fermion is given by
\begin{equation}
\label{massive_cross_section}
 \sigma_\text{abs,1/2,m} = g(R)^{-p/2} A \exp\left(2m\int^{R}_\infty dr\sqrt{g}\right)\,.
\end{equation}
Note that there is no factor of 2 appearing in front compared to~\eqref{sigma_fermion}. For non-vanishing mass $m$, the low-energy absorption cross-section entirely comes from the s-wave and we may neglect the p-wave contribution. For $m=0$, the s- and p-wave contributions to the cross-section are equal and sum up to the given factor of 2 in~\eqref{sigma_fermion} (cf. figure 1 in~\cite{Unruh:1976fm} in the four-dimensional Schwarzschild case). A different way to see this is to notice that the mass terms in~\eqref{massive_Dirac_equation_FG} roughly behave as higher angular momentum modes which distinguish between the $\lambda_n^\pm$ modes, although they are degenerate and contribute equally to $\sigma_\text{abs,1/2}$ for $n=0$ in the massless limit.

\subsection{Application to non-dilatonic black branes}

We now use these results to compute the supersound diffusion constant $D_s$ for a specific class of black brane space-times. In later sections, we will use the direct holographic methods of~\cite{Policastro:2008cx,Kontoudi:2012mu} for the calculation of the same quantity. The results will agree and therefore provide a useful cross-check.

The metrics under consideration are non-dilatonic $p=d-1$ branes in AdS$_{d+1}$~\cite{Horowitz:1998ha}:
\begin{align}
	ds^2 = - f(r) dt^2 + \frac{r^2}{l^2} \sum_{i=1}^{d-1} dx_i^2 + \frac{dr^2}{f(r)}\,,\quad\text{where} \quad f(r)=\frac{r^2}{l^2} -  \frac{R^d}{l^2 r^{d-2}}\,,\label{metric} 
\end{align}
and the AdS radius is $l$. The AdS boundary is at $r\rightarrow \infty$ and the horizon at $r=R$. For $d=4,\,3,\,6$ the space-time represents the near-horizon limit of near-extremal D3-, M2- and M5-branes reducing the sphere. For other dimensions, these can be understood by taking the near-horizon limit of a class of near-extremal $D$-dimensional black $p$-brane solutions~\cite{Gibbons:1994vm,Gibbons:1993sv} which generalize D3-, M2- and M5-branes. The Hawking temperature $T$, Bekenstein-Hawking entropy $S$ and Abbott-Deser mass $M$ of the brane (see e.g.~\cite{Horowitz:1998ha}) are given by
\begin{equation}
 T = \frac{d R}{4 \pi l^2}\,,\quad S=\frac{A}{4G}= \frac{2 \pi}{\kappa_{d+1}^2} \frac{R^{d-1}}{l^{d-1}}V_\parallel\,,\quad M = \frac{d-1}{2 \kappa_{d+1}^2} \frac{R^d}{l^{d+1}}V_\parallel=\epsilon\,V_\parallel\,,\label{TSM}
\end{equation}
where $V_\parallel$ is the volume of the brane measured in the coordinates $x_i$ at constant $t$. 

Evaluating~\eqref{massive_cross_section} upon suitable regularization in the UV in the spacetime~\eqref{metric} one arrives at (note $ml =(d-1)/2$)
\begin{equation}
 \frac{\sigma}{A}=\exp\left[2m\int^{R}_\infty dr\left(\frac{1}{\sqrt{f(r)}}-\frac{l}{r}\right)\right]=\frac{1}{4}\,2^{\frac{2}{d}}\,.\label{sigma_div_A1}
\end{equation}

We may also arrive at this result by evaluating~\eqref{sigma_fermion} in the higher-dimensional black brane space-time before reducing on the sphere. This is due to the fact that the gravitino is massless in the higher-dimensional space-time and gains a ``mass'' $ml =\frac{d-1}{2}$ upon sphere reduction. We are now going to show that this approach also gives the same result~\eqref{sigma_div_A1}. 

The asymptotically flat Gibbons, Horowitz and Townsend non-dilatonic black $p$-branes in $D$ space-time dimensions~\cite{Gibbons:1994vm} can be written as
\begin{equation}
  ds^2=H(r)^{-\frac{2}{p+1}} \left[ - F(r) dt^2 + d\vec{x}_p^2\right]+H(r)^{\frac{2}{D-p-3}}\left[F(r)^{-1}dr^2 + r^2 d\Omega^2_{D-p-2}\right]\label{GHT_black_branes}
\end{equation}
with $H(r)=1 +\left(\frac{l}{r}\right)^{D-p-3}$ and $F(r)=1 -\left(\frac{R}{r}\right)^{D-p-3}$. After some coordinate redefinitions, the near-horizon geometry of these reduces to
\begin{equation}
	ds^2=-f(r)dt^2 + \frac{r^2}{l_\text{AdS}^2} d\vec{x}_p^2+f(r)^{-1} dr^2+l^2_\text{Sph}d\Omega_{D-p-2}^2\,,
\end{equation}
where
\begin{equation}
	f(r)=\frac{r^2}{l_\text{AdS}^2}-\left(\frac{R}{l_\text{AdS}}\right)^2\left(\frac{R}{r}\right)^{p-1}\quad\text{and}\quad l=l_\text{Sph}=\left(\frac{D-p-3}{p+1}\right)l_\text{AdS}\,.
\end{equation}
This is indeed a black $p$-brane in AdS$_{p+2}\times$S$^{D-p-2}$ and clearly, depending on $p$ and the overall dimension $D$, it is either the near-horizon geometry of the selfdual three-brane of ten-dimensional supergravity~\cite{Horowitz:1991cd} or the M2- or M5-brane of eleven-dimensional supergravity~\cite{Duff:1990xz,Gueven:1992hh} or the self-dual string of six-dimensional supergravity~\cite{Duff:1993ye}. Reducing on the sphere gives~\eqref{metric}.

Now, we would like to transform the relevant part of \eqref{GHT_black_branes} into the form \eqref{DGM_black_hole} used by Das, Gibbons and Mathur in their theorem for the low-energy absorption cross section of a spin $1/2$ particle in an (asymptotically flat) black hole background. For this we need to require
\begin{equation}
	F(r)^{-1}dr^2+r^2 d\Omega_{D-p-2}^2\equiv g(\tilde{r})\left[d\tilde{r}^2+\tilde{r}^2d\Omega_{D-p-2}^2\right]\,.
\end{equation}
Solving this for $\tilde{r}$, we directly obtain
\begin{equation}
	\tilde{r}= c\, r \left(1+\sqrt{F(r)}\right)^{\frac{2}{D-p-3}}\,.
\end{equation}
The requirement that we have $r=\tilde{r}$ for $r \rightarrow \infty$ yields $c=2^{-\frac{2}{D-p-3}}$. In the $\tilde{r}$ coordinates the horizon is at $\tilde{R}=c\,R$ with $g(\tilde{R})=c^{-2}$.

Now, we would like to evaluate the absorption cross section for the spin $1/2$ particle~\eqref{sigma_fermion}
\begin{equation}
  \sigma=2 \,g(\tilde{R})^{-\frac{D-p-2}{2}}\,A\,.\label{sigma_fermion2}
\end{equation}
It now easily follows that
\begin{equation}
	\frac{\sigma}{A}=2^{-2+\frac{D-p-5}{D-p-3}}\,.
\end{equation}
This already gives the supersound diffusion constant calculated for the D3-, M2- and M5-branes and may be transformed directly into the result we already obtained previously~\eqref{sigma_div_A1}.

Given a $D$-dimensional action which consists just of non-dilatonic Einstein gravity and the action for a $p$-form field, a necessary condition for the truncation of the massive modes after a sphere reduction to be consistent is given by~\cite{Cvetic:2000dm}
\begin{equation}
	\left(D-p-5\right)\left(p-1\right)=4\,.\label{consistent_truncation}
\end{equation}
Using this we get agreement with~\eqref{sigma_div_A1},
\begin{equation}
	\frac{\sigma}{A}=\frac{1}{4}\,2^{\frac{2}{d}}\,.\label{sigma_div_A2}
\end{equation}
Note however that the condition~\eqref{consistent_truncation} singles out D3-, M2- and M5-brane theories for integers $D$ and $p$. We would have to think along the lines of generalized dimensional reduction~\cite{Kanitscheider:2009as} to allow for the other values. Note that to our knowledge, this has however not yet been worked out for spinors.

We are now in the position to put together our results to obtain an expression for the supersound diffusion constant $D_s$ which may be compared to~\cite{Policastro:2008cx,Kontoudi:2012mu}.

Noting that for the branes~\eqref{metric} with $T$, $A$ and $\epsilon$ given in~\eqref{TSM}, we may use the conformal relation~\eqref{conformal_diffusion_constants} and~\eqref{Kubo_sigma} to get
\begin{equation}
 2 \, \pi T D_s =\frac{2^{2/d} d(d-2)}{2(d-1)^2}\,. \label{supersound_diffusion_general}
\end{equation}
This result agrees with the $d=4$ result of~\cite{Policastro:2008cx,Kontoudi:2012mu} and seems to agree with the $d=3$ result which so far has only been found numerically in~\cite{Gauntlett:2011mf,Gauntlett:2011wm}. Furthermore, the result for $D_s$ vanishes for $d=2$ since (super)gravity in three dimensions has no propagating degrees of freedom.

Note that when using~\eqref{sigma_fermion2}, the boundary diffusion constant is completely determined in terms of \emph{horizon data} only. So, in other words, there seems to be no non-trivial bulk evolution similar to the non-evolution through the bulk of $\eta / s$. In the latter case, the independence on the radial coordinate equates the boundary field theory's result of $\eta / s$ with its membrane paradigm value $1/4\pi$~\cite{Iqbal:2008by}. However, we are dealing with a special coordinate system here, in which the flow seems to be trivial. This generically does not agree with the coordinate system in which $r$ is the field theory's energy scale, but rather appears to be trivial in $\tilde{r}$ where $f(r)^{-1} dr^2 =\frac{d\tilde{r}^2}{\tilde{r}^2}$. It would be interesting to study the setup more intensely along the lines of~\cite{Iqbal:2008by} and~\cite{Iqbal:2009fd}.

\section{Supersound diffusion constant from the transverse gravitino}\label{transverse}
We now turn our focus to the computation of the supersound diffusion constant by extending the holographic computations of~\cite{Policastro:2008cx,Kontoudi:2012mu} for the D3-brane to the case of M2- and M5-brane theories. Simultaneously, we extend it to the aforementioned class of near-extremal non-dilatonic $p=d-1$ branes in AdS$_{d+1}$~\eqref{metric}. The chemical potentials for R-charges are taken to vanish.

We are going to present the calculation via the transversal mode (as in~\cite{Kontoudi:2012mu}), which requires solving a gravitino's equation of motion to $0$'th order in $\omega$ and $k$ and using a Kubo formula. The longitudinal calculation (as in~\cite{Policastro:2008cx}) is technically more difficult, since it also requires solving the equations of motion to linear order in $\omega$ and $k$, and will be covered in the subsequent section.

The bulk action for the linearized gravitino is given by
\begin{equation}
 S\propto\int d^{d+1}x\sqrt{-g}\bar{\Psi}_\mu \left(\Gamma^{\mu\nu\rho}D_\nu-m\Gamma^{\mu\rho}\right)\Psi_\rho\,,\label{bulk}
\end{equation}
where the normalization will first be unimportant.

The covariant derivative acts on spinors as $D_\mu = \partial_\mu + \frac{1}{4}\omega_\mu^{ab}\gamma_{ab}$ where the only non-vanishing components of the spin-connection for the background~\eqref{metric} are $\omega_{t}^{0d}=\frac{1}{2} f^\prime$ and $\omega_{x_i}^{di}=-\frac{\sqrt{f}}{l}$ for each $i=1,\ldots, d-1$. Furthermore, $ml =\frac{d-1}{2}$ is required for linearized supergravity to hold in AdS$_{d+1}
$~\cite{Townsend:1977qa} and ensures the gravitino to have the correct degrees of freedom of a massless gravitino field~\cite{Deser:1977uq}.

Using the standard gauge condition for the gravitino $\Gamma^\mu \Psi_\mu=0$ where $\mu=0,\ldots,d$ (which also implies $D^\mu \Psi_\mu=0$) the Rarita-Schwinger equation may be simplified to 
\begin{equation}
 \left(\slashed{D}+m\right) \Psi_\mu =0\,.\label{eom1}
\end{equation}

These are complicated equations of motion which couple different components of the gravitino\footnote{From here on, all vector-like indices are Lorentz indices, implicitly using appropriate vielbeins.}. However, assuming a boundary space-time dependence $e^{-i \omega t + i k x_1}$ of the gravitino, we may use the projector~\eqref{transverse_projector} to project to the $k^\mu$-transverse components of the gravitino which have spin $3/2$ under the transverse $O(d-2)$ that preserves the boundary wave vector. These components $\eta_i =\Psi_i - \frac{1}{d-2}\gamma_i \gamma^j \Psi_j$ ($i,j\neq 1$) decouple, so that the equation of motion for one of them in the background~\eqref{metric}, call it $\eta$, reads:
\begin{equation}
 \eta^\prime -\frac{i \omega}{f} \gamma^d \gamma^0 \eta + \frac{i k l}{r \sqrt{f}} \gamma^d\gamma^1 \eta +\frac{f^\prime}{4 f} \eta + \frac{d-1}{2r}\eta - \frac{m}{\sqrt{f}}\gamma^d \eta =0\,.\label{eom_eta}
\end{equation}

Note that this is exactly the equation of motion for a minimally coupled spin 1/2 fermion of mass $m$ in the given space-time as was used heavily in the previous section.

We may now expand $\eta$ in a basis of eigenspinors of $\gamma^d$ and $i\gamma^{1}\gamma^2$ since these commute with each other. Let the basis spinors be given by
\begin{align}
 \gamma^d a^\pm = \pm a^\pm\,,\quad  \gamma^d b^\pm = \pm b^\pm\,,\quad
 i\gamma^{1}\gamma^2 a^\pm =+ a^\pm\,,\quad  i\gamma^{1}\gamma^2 b^\pm = - b^\pm\,.\label{basis_spinors}
\end{align}
Then one can write
\begin{align}
\eta = \eta^{a+} a^+ + \eta^{a-} a^- +\eta^{b+} b^++\eta^{b-} b^-\,.\label{eigenspinor_decomposition}
\end{align}
Note that from the $d$-dimensional point of view, $\eta^\pm= \frac{1}{2} \left(1\pm \gamma^d\right)\eta$ are Weyl spinors of opposite chirality when $d$ is even. For $d$ odd, both are $d$-dimensional Dirac spinors. We may additionally choose a particular set of $\gamma$ matrices such that 
\begin{align}
\gamma^0 a^\pm=\pm a^\mp\,,\quad\gamma^0 b^\pm = \mp b^\mp\,,\quad\gamma^1 a^\pm = \pm i \, b^\mp\,,\quad\gamma^1 b^\pm = \pm i\, a^\mp\,,
\end{align}
which is all compatible with the Dirac algebra $\left\{\gamma^\mu,\gamma^\nu\right\}= 2 \eta^{\mu\nu}$, to reduce~\eqref{eom_eta} to four equations for the spinors $\eta^{a\pm}$ and $\eta^{b\pm}$.

To get the retarded real-time correlator we need to solve~\eqref{eom_eta} using ingoing boundary conditions at the horizon~\cite{Son:2002sd,Policastro:2002se}. For imposing these on our solutions, we need to look at the most singular part of~\eqref{eom_eta} close to the horizon, where we may restrict ourselves to the $+$ eigenspace with respect to $i\gamma^{1}\gamma^2$ (the $-$ eigenspace goes analogously). Close to the horizon the solutions $\eta^{a\pm}$ are thus required to behave as
\begin{equation}
  \eta^{a\pm}\sim \left(r-R\right)^{-\frac{1}{4} - \frac{i\omega}{4\pi T}}\eta^{a\pm}_{0}\,,\quad \eta^{a+}_{0}=\eta^{a-}_{0}\,.\label{sol_hor}
\end{equation}

Since we are going to use the Kubo formulae~\eqref{Kubo}~\eqref{Kubo2} to determine the supersound diffusion constant later on, in which we need to take the limit of small frequencies and momenta, we may directly set $\omega=k=0$ in~\eqref{eom_eta} and straightforwardly integrate the equations for components $\eta^\pm$,
\begin{equation}
 \eta^\pm =c^\pm f^{-1/4} r^{-\frac{d-1}{2}} \left(r^{d/2}+\sqrt{r^d-R^d}\right)^{\pm \frac{d-1}{d}}\,.\label{sol}
\end{equation}
Using~\eqref{sol_hor}, we may derive the relation $c^+ =R^{1-d}c^-$ between the integration constants.

At the boundary, the given solutions~\eqref{sol} have the asymptotic behaviour
\begin{align}
  \eta^+&\sim c^+ l^{1/2} \left(\frac{2}{2^{1/d}}\right) r^{-1/2}\equiv \phi\, r^{-1/2}\,,\\
  \eta^-&\sim c^- l^{1/2} \left(\frac{2^{1/d}}{2}\right) r^{1/2-d}\equiv \chi\, r^{1/2-d}\,.
\end{align}
Clearly, the first term is a source term which couples to an operator in the boundary conformal field theory of conformal dimension $\Delta=\frac{1}{2}(d+2|m l|)$~\cite{Corley:1998qg,Volovich:1998tj} and the second is related to the operator's expectation value. At the boundary we further find $\chi = R^{d-1} 2^{2/d-2} \phi$.

We now insert this asymptotic behaviour into the boundary term of the gravitino~\cite{Corley:1998qg,Volovich:1998tj} (for spin 1/2 fermions see~\cite{Henningson:1998cd,Mueck:1998iz}) to compute the Green's function of the dual operators. We do not need to worry about holographic renormalization~\cite{deHaro:2000xn} here, since in the limit $\omega,k\rightarrow 0$ there are no divergences; the first covariant counterterm has to be inserted at order $O(k^\mu)$~\cite{Kontoudi:2012mu}. We have
\begin{equation}
 S_{\text{bdy}}=\frac{4}{2 \kappa_{d+1}^2}\int d^d x \sqrt{-h}\, \bar{\Psi}_i h^{ij}\Psi_j \, . \label{bdy}
\end{equation}

The normalization of this boundary term was fixed in~\cite{Kontoudi:2012mu} (for the AdS$_5$ case) by using the $T=0$ superspace correlators in the boundary field theory. Here we proceed analogously, which will also explain the non-standard kinetic term chosen in~\eqref{non_canonical_bulk_fermion}. The supersymmetry algebra in $d$ space-time dimensions without central extensions has the following form~\cite{Strathdee:1986jr}
\begin{align}
  \left\{Q,\bar{Q}\right\}=2 \gamma^\mu P_\mu\,,
\end{align}
with appropriate chiral projections $\frac{1}{2}\left(1\pm \gamma^d\right)$ applied in even dimensions, e.g.\ giving $\left\{Q_\alpha,\bar{Q}_{\dot{\beta}}\right\}=2 \sigma_{\alpha \dot{\beta}}^\mu P_\mu$ in 4 dimensions for Weyl spinors $Q_\alpha$ and $\bar{Q}_{\dot{\beta}}$.

The commutation relations extend to the supercurrent multiplet which includes the energy momentum tensor and the supersymmetry current. The same applies to the R-symmetry current whose contributions will not be of importance for our argument and which we therefore suppress in the following. The commutation relations are given by
\begin{align}
  \label{SUSY_algebra_T}
  \left\{Q,\bar{S}_\mu\right\}&=2 \gamma^\nu T_{\mu\nu}\,,\\
  \label{SUSY_algebra_S}
  \left[\bar{Q},T_{\mu\nu}\right]&=-\frac{i}{8}\partial^\rho \bar{S}_\mu\left(\gamma_\nu \gamma_\rho - \gamma_\rho \gamma_\nu\right)-\frac{i}{8}\partial^\rho \bar{S}_\nu\left(\gamma_\mu \gamma_\rho - \gamma_\rho \gamma_\mu\right)\,,
\end{align}
again with chiral projectors implicitly assumed in even dimensions (for four dimensions see~\cite{Sohnius:1985qm}). We may use these to relate the two-point function of the energy-momentum tensor~\cite{Liu:1998bu,Arutyunov:1999nw} to the two-point function of the supersymmetry current~\cite{Corley:1998qg}. The former is given by
\begin{equation}
   \left\langle T_{\mu\nu}(x) T_{\rho\sigma}(0)\right\rangle = \frac{C_d}{x^{2d}}\mathcal{I}_{\mu\nu,\rho\sigma}(x)\,,
\end{equation}
where the normalization in $d$ dimensions is $C_d=\frac{1}{2 \kappa_{d+1}^2}\frac{2 d(d+1)}{(d-1)}\frac{\Gamma(d)}{ \pi^{d/2}\Gamma(d/2)}$. Furthermore, for $J_{\rho\nu}(x)=\left(\eta_{\rho\nu}-2\frac{x_\rho x_\nu}{x^2}\right)$ the tensor $\mathcal{I}_{\mu\nu,\rho\sigma}(x)=\frac{1}{2}\left(J_{\mu\rho}J_{\nu\sigma}+J_{\mu\sigma}J_{\nu\rho}\right)-\frac{1}{d}\eta_{\mu\nu}\eta_{\rho\sigma}$ represents the inversion $x_\mu \rightarrow \frac{x_\mu}{x^2}$ on symmetric traceless tensors~\cite{Erdmenger:1996yc}.

The supersymmetry current correlator on the other hand can be written as
\begin{equation}
   \left\langle S_{\mu}^+(x)\bar{S}^-_\nu(0)\right\rangle = 2 a \left(\frac{\kappa^2_{d+1}}{d+1}C_d\right)\left(\delta_\mu^\rho-\frac{1}{d}\gamma_\mu\gamma^\rho\right) \frac{\gamma^\sigma x_\sigma}{x^{2d}}J_{\rho\nu}(x)\,,
\end{equation}
where $S_\mu^\pm = \frac{1}{2}(1\pm \gamma^d)S_\mu$ and $a$ represents the normalization of the boundary action $S_{\text{bdy}}=a \int d^d x \sqrt{-h}\, \bar{\Psi}_i h^{ij}\Psi_j$ we would like to determine. Using the given part of the higher-dimensional supersymmetry algebra~\eqref{SUSY_algebra_T}~\eqref{SUSY_algebra_S} and the fact that in a supersymmetric ground-state $Q |0\rangle=\bar{Q} |0\rangle=0$, we may relate the two correlators to one another and by this determine the normalization to be\footnote{In~\cite{Kontoudi:2012mu} the normalization is stated as $\mathcal{N}=\frac{N_c^2}{\pi^2}=\frac{8}{2\kappa_{5}^2}$ for the AdS radius set to $l=1$. This different factor of 2 compared to~\eqref{boundary_term_normalization} for $d=4$ seems to be compensated by the fact that $\eta_2 =\Psi_2 - \frac{1}{2}\gamma_2\left(\gamma^2\Psi_2+\gamma^3\Psi_3\right)$ and $\eta_3 =\Psi_3 - \frac{1}{2}\gamma_3\left(\gamma^2\Psi_2+\gamma^3\Psi_3\right)$ in $d=4$ are essentially the same by construction due to the vanishing 
spin 1/2 part identity $\gamma^2 \eta_2 +\gamma^3\eta_3 =0$. Therefore, our results in the end nevertheless agree with~\cite{Kontoudi:2012mu}.}
\begin{equation}
   a=\frac{4}{2\kappa^2_{d+1}}\, .\label{boundary_term_normalization}
\end{equation}
This is essentially independent of the dimension and the possible Weyl character of the supercharges in even dimensions. Having fixed the boundary term normalization, the normalization of the kinetic term is not free any more although it plays no role in the solution to the equation of motion. As in the case of the Gibbons-Hawking boundary term, we may also argue the gravitino boundary term~\eqref{bdy} to be present for a well defined variational principle. This imposes a relative normalization between~\eqref{bdy} and~\eqref{bulk}~\cite{Henneaux:1998ch,Rashkov:1999ji} which results in the specific normalization used in~\eqref{non_canonical_bulk_fermion} for one of the transverse gravitini.

With this normalization, we may now finally calculate the retarded Green's function of the transverse supersymmetry current operator, which is dual to $\eta$,
\begin{equation}
 G^R =\frac{4 i}{2 \kappa_{d+1}^2} \left(\frac{R}{l}\right)^{d-1}  \textrm{diag}\left(\frac{2^{2/d}}{4},\ldots,\frac{2^{2/d}}{4}\right)\,.
\end{equation}
We now proceed by using the $d$-dimensional Kubo formula~\eqref{Kubo}~\eqref{Kubo2} for the supersound diffusion constant to get
\begin{align}
 {\epsilon} D_{3/2} &=\frac{1}{\kappa_{d+1}^2} \left(\frac{R}{l}\right)^{d-1} \frac{2^{2/d}}{2}\,.
\end{align}
Note that in the end formula it does not matter if we imposed a Weyl constraint in even dimensions. If so, the Green's function would have half as many entries but this would be compensated by an additional factor of two in the Kubo formula~\eqref{Kubo}.

Using the relation between $D_s$ and $D_{3/2}$~\eqref{conformal_diffusion_constants}, the field theory's equilibrium energy density ${\epsilon} = M / V_\parallel$ and temperature $T$~\eqref{TSM}, we arrive at
\begin{equation}
 2 \pi T D_s = \frac{2^{2/d} d (d-2)}{2 (d-1)^2}\,,\label{supersound_diffusion}
\end{equation}
which agrees with~\eqref{supersound_diffusion_general}.

\section{Supersound diffusion constant from the phonino pole}\label{longitudinal}

We now determine the phonino dispersion relation 
\begin{equation}
  \omega = v_s k - i D_s k^2
\end{equation}
from the pole of the longitudinal supersymmetry current correlator, closely following and generalizing the computation of~\cite{Policastro:2008cx} to $d$ dimensions. This not only reproduces the already established result for the diffusion constant $D_s$~\eqref{supersound_diffusion_general}~\eqref{supersound_diffusion} by a further computation, but also adds additional confidence to the result since, along the way, it determines the value of the supersound velocity $v_s$, which agrees with the one dictated by conformal invariance $v_s=\frac{P}{\epsilon}=\frac{1}{d-1}$.

For determining the dispersion relation we need to solve part of the full set of the gauge-fixed equations of motion $(\slashed{D}+m)\Psi_\mu=0$ on the gravity side to linear order in $\omega$ and $k$, using ingoing boundary conditions at the horizon. We then need to read off the source terms of the dual CFT operators in the expansion data at the AdS boundary from which we can then easily extract the pole of the supersymmetry current correlator. 

Again we assume our gravitino to be a Dirac vector-spinor. Imposing e.g.\ an additional Majorana constraint in certain dimensions would basically just restrict the integration constants we encounter in the following to be real. This has no influence on the pole structure so is irrelevant for our argument.

The equations of motion are given by (again, all vector-like indices are Lorentz indices from now on, after transforming with appropriate vielbeins)
\begin{subequations}
\label{eom2}
\begin{align}
	\gamma^d \Psi_0^\prime - \frac{i \omega}{f} \gamma^0 \Psi_0 - \frac{f^\prime}{2 f} \gamma^0 \Psi_d +\frac{f^\prime}{4 f} \gamma^d \Psi_0 + \frac{i k l}{r\sqrt{f}} \gamma^1 \Psi_0 + \frac{d-1}{2 r} \gamma^d \Psi_0 + \frac{m}{\sqrt{f}}\Psi_0 =0 \,,\\	
	\gamma^d \Psi_d^\prime - \frac{i \omega}{f} \gamma^0 \Psi_d - \frac{f^\prime}{2 f} \gamma^0 \Psi_0 +\frac{f^\prime}{4 f} \gamma^d \Psi_d + \frac{i k l}{r\sqrt{f}} \gamma^1 \Psi_d + \frac{1}{r} \left(\frac{d+1}{2} \gamma^d \Psi_d+\gamma^0 \Psi_0\right) + \frac{m}{\sqrt{f}}\Psi_d =0\,,\\
	\gamma^d \Psi_j^\prime - \frac{i \omega}{f} \gamma^0 \Psi_j +\frac{f^\prime}{4 f} \gamma^d \Psi_j + \frac{i k l}{r\sqrt{f}} \gamma^1 \Psi_j + \frac{1}{r} \gamma^j \Psi_d+ \frac{d-1}{2r} \gamma^d \Psi_j + \frac{m}{\sqrt{f}}\Psi_j =0\,,
\end{align}
\end{subequations}
for $j=1,\ldots, d-1$, where we will use that the massless gravitino in $AdS_{d+1}$ has $m l =\frac{d-1}{2}$ as argued e.g.\ in~\cite{Gauntlett:2011mf,Gauntlett:2011wm} following~\cite{Grassi:2000dm}. A relation, which will turn out very useful for decoupling the different components of~\eqref{eom2} is derived by using the gauge condition $\gamma^\mu \Psi_\mu=0$, taking its radial derivative and using the equations of motion given above. The calculation then gives the following constraint equation
\begin{equation}
\left(\frac{f^\prime}{2 f} \gamma^d - \frac{2 i \omega}{f} \gamma^0 + \frac{2 i k l}{r \sqrt{f}} \gamma^1 - \frac{2 m}{\sqrt{f}}+\frac{d-2}{r}\gamma^d\right)\gamma^d \Psi_d + \frac{2 i k l}{r \sqrt{f}} \Psi_1+ \left(\frac{f^\prime}{2 f} \gamma^d - \frac{2 i \omega}{f} \gamma^0  -\frac{1}{r}\gamma^d\right)\gamma^0 \Psi_0  = 0\,,
\label{constraint_equation}
\end{equation}
which we will make frequent use of throughout the calculation.\\

We now start to solve the equations of motion in the hydrodynamical limit. Therefore, we expand the gravitino to first order in $\omega$ and $k$ and solve perturbatively in these quantities:
\begin{equation}
	\Psi_\mu = \psi_\mu + \omega\varphi_\mu  + k\chi_\mu \,.
\end{equation}
Let us start with the lowest order terms, where we can basically set $\omega=k=0$ in~\eqref{eom2} and~\eqref{constraint_equation}. At this order the equation for $\Psi_d$ is diagonal:
\begin{equation}
	\psi_d^\prime +\left( \frac{3 f^\prime}{4 f} + \frac{3(d-1)}{2r} - \frac{m}{\sqrt{f}}\gamma^d\right) \psi_d =0\,.
\end{equation}
Similarly, we get
\begin{align}
	\psi_0^\prime +\left( \frac{ f^\prime}{4 f} + \frac{d-1}{2r} + \frac{m}{\sqrt{f}} \gamma^d\right) \psi_0 &= -\frac{f^\prime}{2 f} \gamma^0 \gamma^d \psi_d\,,\\
	\psi_1^\prime +\left( \frac{ f^\prime}{4 f} + \frac{d-1}{2r} + \frac{m}{\sqrt{f}} \gamma^d \right) \psi_1 &= \frac{1}{r} \gamma^1 \gamma^d \psi_d\,.
\end{align}
The equation for $\psi_d$ can be integrated directly after decomposing it analogously to~\eqref{eigenspinor_decomposition} while the other ones can be solved by the method of integrating factors given the solution for $\psi_d$ and the action of $\gamma^0$ and $\gamma^1$ on the eigenspinors $a^\pm$ and $b^\pm$ given earlier~\eqref{basis_spinors}. Let us denote integration constants by $a_i, b_i, c_i, d_i$ when integrating the zero'th order $a^\pm$, $b^\pm$ parts of the gravitino component $\psi_i$. In the following we use the notation $\psi_\mu=(\psi_\mu^{a+}\,, \psi_\mu^{a-}\,,\psi_\mu^{b+} \,,\psi_\mu^{b-} )^T$.

Near the horizon we find up to $O(r-R)^{-1/4}$
\begin{align}
	\psi_d=\frac{l^{3/2}R^{d-3}}{(dR)^{3/4} }\begin{pmatrix} a_d \\ a_d \\ c_d \\ -c_d \end{pmatrix}  (r-R)^{-3/4} \,,
\end{align}
where we have already imposed ingoing boundary conditions $\propto \left(r-R\right)^{-\frac{i \omega}{4 \pi T}}$ at the horizon which translate into $a_d = b_d$ and $c_d = - d_d$. It will be exactly this condition for most of the other functions at all orders in $\omega$ and $k$, so we will not explicitly state the near-horizon analysis any more. Similar considerations for example yield
\begin{align}
	\psi_0=\frac{l^{3/2}R^{3-d}}{(dR)^{3/4} }\begin{pmatrix} a_d \\ a_d \\ c_d \\ -c_d \end{pmatrix}  (r-R)^{-3/4}\,,
\end{align}
\begin{align}
	\psi_1=-i\,\frac{\sqrt{l} \,R^{-1/4-d}}{d^{5/4}\,(d-1)}\begin{pmatrix} 2\,(d-1)\,l\, R^2\, c_d-d\, R^d\, c_\Sigma \\ 2\,(d-1)\,l\, R^2 \,c_d-d\, R^d\, c_\Sigma \\ -2\,(d-1)\,l\, R^2 \,a_d+d\, R^d\, a_\Sigma \\ 2\,(d-1)\,l\, R^2\, a_d-d\, R^d\, a_\Sigma \end{pmatrix}  (r-R)^{-1/4}\,,
\end{align}
where the matching of integration constants in $\psi_d$ and $\psi_0$ is due to the constraint equation~\eqref{constraint_equation}, and the integration constants $a_1,c_1$ for $\psi_1$ are written in a way to match a convenient notation that will be explained in more detail below.

The interesting AdS boundary behaviour is given by 
\begin{align}
\begin{pmatrix} \psi_0^{a-} \\ \psi_0^{b-} \\ \psi_1^{a-} \\ \psi_1^{b-}\end{pmatrix}
=\frac{2^{(d-1)/d}\,i\, \sqrt{l}}{d-1}  \begin{pmatrix}  0 \\ 0\\ c_\Sigma \\ a_\Sigma \end{pmatrix} r^{-1/2}\,,
\label{lowest_order_boundary_solution}
\end{align}
where we have only shown the source terms.

So, indeed, we do find sources for our CFT operators. In particular, it is interesting to see that there are no source terms in the time component of the gravitino, only in the longitudinal ones.\\

We must now proceed to the terms linear in $\omega$ and $k$. The strategy is exactly the same as before. We decouple the equations of motion and integrate them using the solutions given earlier by the method of integrating factors. We then obtain solutions on which we must impose ingoing boundary conditions at the horizon. Then we read off the source terms near the boundary. Although conceptionally this does not pose problems any more and it is indeed possible to decouple the equations up to a non-homogeneous, explicitly known part, the actual expressions for the integrated solutions become very complicated.

We now introduce another object which will be very handy in the actual computation of the source terms. Let us define
\begin{align}
	\Psi_{\Sigma} =\left(d-2\right)\gamma^1 \Psi_1 - \gamma^2 \Psi_2-\ldots-\gamma^{d-1}\Psi_{d-1}\,.
\end{align}
Looking at the equations of motion~\eqref{eom2}, we see that its equation of motion partly decouples from the other gravitino components:
\begin{equation}
	\Psi_\Sigma^\prime+\frac{i\omega}{f}\gamma^0\gamma^d\Psi_\Sigma+\frac{f^\prime}{4 f} \Psi_\Sigma -\frac{i k l}{r\sqrt{f}}\gamma^d\left(2(d-2)\Psi_1-\gamma^1\Psi_\Sigma\right)+\frac{(d-1)}{2r} \Psi_\Sigma-\frac{m}{\sqrt{f}}\gamma^d \Psi_\Sigma=0
\end{equation}
We may use the solution to this field together with the gauge condition $\gamma^\mu\Psi_\mu=0$ to solve for $\Psi_1$ which contains our longitudinal source terms,
\begin{equation}
	\Psi_1 =\frac{1}{(d-1)}\left(\gamma^1 \Psi_\Sigma -\gamma^1\gamma^0\Psi_0 -\gamma^1 \gamma^d \Psi_d\right)\,.
	\label{calculation_of_phi_2}
\end{equation}

These steps seem to be necessary computationally since the direct analytic integration of the equations of motion for $\varphi_1$ and $\chi_1$ appears to be far too complicated even for computer algebra.

Let us begin with the calculation to first order in $\omega$. What do the relevant equations of motion for $\varphi_\mu$ now look like? They are given by
\begin{align}
	\varphi_d^\prime +\left(\frac{3 f^\prime}{4f}+\frac{3(d-1)}{2r}-\frac{m}{\sqrt{f}} \gamma^d \right)\varphi_d &=\frac{i}{f} \gamma^0\gamma^d \psi_d-\frac{2i}{f}\psi_0\,,\\
	\varphi_\Sigma^\prime+\left(\frac{f^\prime}{4 f}  +\frac{d-1}{2r} -\frac{m}{\sqrt{f}}\gamma^d \right)\varphi_\Sigma &=-\frac{i}{f}\gamma^0\gamma^d\psi_\Sigma\,,
\end{align}
which can straightforwardly be solved by the method of integrating factors since the solutions for $\psi_d,\psi_0$ and $\psi_\Sigma$ are known. Imposing ingoing boundary conditions at the horizon then gives a similar structure on the integration constants as before and we can read off the source terms after calculating $\varphi_0$ from~\eqref{constraint_equation} and $\varphi_1$ from~\eqref{calculation_of_phi_2}. They are given by
\begin{align}
\begin{pmatrix} \varphi_0^{a-} \\ \varphi_0^{b-} \\ \varphi_1^{a-} \\ \varphi_1^{b-}\end{pmatrix}
=-\frac{ 2^{(3d+1)/d}\, l^{7/2} \, R^{1-d}}{d^2 }\begin{pmatrix}  -i \, (d-1)\,a_1 \\   i \, (d-1)\,c_1\\  c_1\\  a_1  \end{pmatrix}  r^{-1/2}\,.
\end{align}

Now, the calculations for $\chi_\mu$ are in spirit similar to the ones before. Computationally however, they are more difficult. As before, we will only give the starting point and the results. The equations of motion, which have to be solved are
\begin{align}
	\chi_d +\left(\frac{3 f^\prime}{4f}+\frac{3(d-1)}{2r}-\frac{m}{\sqrt{f}} \gamma^d \right)\chi_d &=-\frac{i l}{r \sqrt{f}}\left(2\psi_1 +\gamma^1\gamma^d\psi_d\right)\,,\\
	\chi_\Sigma^\prime+\left(\frac{f^\prime}{4 f}  +\frac{d-1}{2r} -\frac{m}{\sqrt{f}}\gamma^d\right) \chi_\Sigma &=\frac{i l}{r\sqrt{f}} \gamma^d\left(-\gamma^1\psi_\Sigma+2(d-2)\psi_1\right)\,.
\end{align}
One more time, we solve by using integrating factors, impose ingoing boundary conditions, calculate $\chi_0$ and $\chi_1$ using~\eqref{constraint_equation} and~\eqref{calculation_of_phi_2} to finally obtain
\begin{align}
\begin{pmatrix} \chi_0^{a-} \\ \chi_0^{b-} \\ \chi_1^{a-} \\ \chi_1^{b-}\end{pmatrix}
=\frac{2^{(d+1)/d}\,l^{5/2} \,R^{1-d}}{(d-1)\,d^2}\begin{pmatrix}
   4\,  (d-1) \,l \, c_d- d\,
   R^{d-2} \,c_\Sigma\\
  4 \, (d-1)\, l \,a_d- d\,
   R^{d-2}\, a_\Sigma\\
-4\,i\,   (d-1)^2 \, l  \, a_d\\
4\,i\, (d-1)^2 \, l \, c_d
\end{pmatrix}  r^{-1/2}\,.
\end{align}
In fact, we also get source terms involving additional $a_\Sigma$ and $c_\Sigma$ dependent terms and new integration constants. When summing up all contributions to get the source terms of the full $\Psi_0$ and $\Psi_1$, we can, as noted in Policastro's paper~\cite{Policastro:2008cx}, redefine our integration constants to have $\omega$ and $k$ dependent terms. So basically, we rename $a_\Sigma\rightarrow a_\Sigma^\prime=a_\Sigma + \alpha\,\omega + \beta\, k$ and similarly for $c_\Sigma$ since these are the integration constants at lowest order~\eqref{lowest_order_boundary_solution}. We have already implicitly done this redefinition in the terms above and therefore only given the really relevant part of the solution.

In total, we now obtain a linear relation between $\Psi_0$, $\Psi_1$ and their boundary values of the form
\begin{align}
\begin{pmatrix} \Psi_0^{a-} \\ \Psi_0^{b-} \\ \Psi_1^{a-} \\ \Psi_1^{b-}\end{pmatrix}
=\mathcal{M} \begin{pmatrix} a_d \\ a_\Sigma \\ c_d \\ c_\Sigma \end{pmatrix}  r^{-1/2}\,,
\end{align}
for a matrix $\mathcal{M}$ that is determined by the source terms computed in the previous two subsections. We now plan to look for non-trivial solutions to this relation, where the boundary values have poles. These poles will then show up in the CFT Green's functions and therefore may be interpreted as the phonino poles~\cite{Kratzert:2003cr}. We have to compute the determinant of $\mathcal{M}$, substituting the relation
\begin{equation}
	\omega = v_s k - i D_s k^2\,.
\end{equation}
We then set the determinant to zero, and solve for $v_s$ and $D_s$, to finally get
\begin{equation}
	v_s=\frac{1}{d-1}\,,\quad\quad 2 \pi T D_s = \frac{2^{2/d} d (d-2)}{2(d-1)^2}\,.\label{supersound_diffusion3}
\end{equation}
The value of the supersound velocity is the one expected from conformal invariance in $d$ dimensions (square of the normal sound velocity). There is no imaginary part of $D_s$ appearing. Of course, the four dimensional result, $v_s=\frac{1}{3}$ and $ 2 \pi T D_s = \frac{4}{9} \sqrt{2}$, is exactly reproduced as for $d=4$ our computation is just a slightly different coordinate version of~\cite{Policastro:2008cx}. It furthermore exactly agrees with~\eqref{supersound_diffusion_general} and~\eqref{supersound_diffusion}.

\section{Generalized dimensional reduction and Dp-branes}\label{Dp_branes}

Before, we have derived the supersound velocity and diffusion constant for the case of asymptotically AdS black branes~\eqref{metric}. However, it is not directly clear how these are related to a string or M theory background. Of course, their boundary field theories are classically conformal as, for instance, can be seen from the dimension dependence of the supersound velocity~\eqref{supersound_diffusion3}
\begin{equation}
v_s=\frac{P}{\epsilon}=\frac{1}{d-1}\quad\Leftrightarrow\quad T^\mu_\mu = - \epsilon + (d-1) P = 0\,.
\end{equation}
However only some of the standard $p$-brane backgrounds of the low-energy limit of type II supergravity and M theory, namely near-horizon D3-, M2- and M5-branes, have boundary CFTs. Since for some of the other D$p$-brane backgrounds the near-horizon limit is also a decoupling limit~\cite{Itzhaki:1998dd} and holographic renormalization has also been established for these~\cite{Wiseman:2008qa,Kanitscheider:2008kd}, we may wonder if it is also possible to derive the supersound diffusion constants for these backgrounds. Actually, from the conformal backgrounds~\eqref{metric} we may already learn a lot about the hydrodynamics of the field theories dual to D$p$-brane backgrounds via what has been called `generalized dimensional reduction'~\cite{Kanitscheider:2009as,Gouteraux:2011qh}. The idea is the following:

Given the standard background of the near-horizon limit of a near-extremal $p$-brane relevant for string theory~\cite{Horowitz:1991cd}, say in string frame, one can perform a Weyl transformation to the so-called dual frame~\cite{Boonstra:1998mp}. In this frame, for $p\neq5$ the metric becomes $AdS_{p+2} \times S^{8-p}$. However, the transformation comes with the price of a non-trivial coupling of the (running) dilaton to the Einstein-Hilbert term of the action. When reducing on the sphere, we have
\begin{equation}
S=-L\int d^{d+1}x\, \sqrt{g}\, e^{\gamma \phi}\left[R+\beta \left(\partial \phi\right)^2+C\right]
\end{equation}
for some constants $L,\gamma,\beta,C$ which depend on the worldvolume dimension of the $p$-brane (for concrete expressions see~\cite{Kanitscheider:2008kd}). This action may also be obtained by dimensionally reducing $(2\sigma +1)$-dimensional pure Einstein gravity with cosmological constant on a $(2\sigma - d)$-dimensional torus. In this reduction the dimensions are formally given in terms of a quantity $\sigma$ which is defined as 
\begin{equation}
\sigma = \frac{d}{2} -\frac{(p-3)^2}{2(p-5)}\,.\label{non_integer_sigma}
\end{equation}
Note that e.g.\ for $p=3$, $\sigma$ is half-integer and so the `higher-dimensional' pure gravity theory just has dimension five, Einstein gravity on AdS$_5$. For $p\neq 3$ however, $\sigma$ is not necessarily half-integer any more. It has been shown in~\cite{Kanitscheider:2008kd} however that holographic renormalization does also make sense for arbitrary, in particular non-half-integer values of $\sigma$! From the point of view of the lower-dimensional theory's equations of motion, this is just a parameter which happens to have some meaning for the special cases of D$p$-branes. Therefore, the transformation to asymptotically AdS spaces, consistent dimensional reduction and analytic continuation in $\sigma$ allow determining the hydrodynamics of non-conformal branes completely by just formally looking at conformal hydrodynamics in higher dimensional AdS theories~\cite{Kanitscheider:2008kd}.

This is what we will exploit now to determine the supersound velocity $v_s$ and diffusion constant $D_s$ for the non-conformal D$p$-branes. Since $v_s=\frac{P}{\epsilon}$, its value is already given in~\cite{Kanitscheider:2009as}:
\begin{equation}
   v_s=\frac{1}{2\sigma-1}
\end{equation}
with $\sigma$ given in~\eqref{non_integer_sigma}. For $D_s$, we should, however, take the upcoming results with some care since for fermions the generalized dimensional reduction has so far not been worked out. We will nevertheless give some support in favour of this procedure.

The black brane space-times given earlier~\eqref{metric} are exactly the interesting class of solutions to Einstein's equations for the higher dimensional theory which is reduced on some torus. We can thus rewrite our results obtained earlier~\eqref{supersound_diffusion_general}~\eqref{supersound_diffusion}~\eqref{supersound_diffusion3} by setting $d=2\sigma$ (so that the whole bulk dimension is $2\sigma+1 = d+1$). We then express $\sigma$ in terms of the dimension of the $p$-brane~\eqref{non_integer_sigma} and get
\begin{equation}
2\pi T D_s = 4\times 2^{\frac{5-p}{7-p}} \frac{(7-p)}{(9-p)^2}\,. \label{Dp_brane_diffusion_constant}
\end{equation}

One may also apply the methods of section~\ref{cross_sections}: Given the near-horizon Einstein-frame metric of a D$p$-brane (see e.g.~\cite{Kovtun:2003wp}) with $f(r)=1-\left(\frac{R}{r}\right)^{7-p}$
\begin{equation}
   ds^2_E=\left(\frac{r}{l}\right)^{\frac{(7-p)^2}{8}}\left(-f(r)dt^2 + dx_1^2+\ldots+dx_p^2\right)+\left(\frac{l}{r}\right)^{\frac{(7-p)(p+1)}{8}}\left(\frac{dr^2}{f(r)}+r^2 d\Omega_{8-p}^2\right)\label{metric_Dp_brane}
\end{equation}
we may again use~\eqref{sigma_fermion} after transforming the metric~\eqref{metric_Dp_brane}, which involves a $(8-p)$-sphere, into the form of~\eqref{DGM_black_hole} to arrive at
\begin{equation}
   \frac{\sigma}{A}=\frac{1}{4}\,2^{\frac{5-p}{7-p}}\,.
\end{equation}
This agrees with what one would get from taking~\eqref{sigma_div_A1} or~\eqref{sigma_div_A2} and applying the generalized dimensional reduction procedure explained above. The energy densities which appear in the Kubo formulae also behave according to generalized dimensional reduction~\cite{Kanitscheider:2008kd,Kanitscheider:2009as}. Therefore, we trust our naive result~\eqref{Dp_brane_diffusion_constant} although, as we again emphasize, generalized dimensional reduction has not been properly worked out for fermions.

Of course, it would be interesting to independently check these D$p$-brane results along the lines of sections~\ref{transverse} and~\ref{longitudinal}.

\section{Conclusion}\label{conclusion}

In this work, we have explicitly computed the supersound diffusion constant $D_s$ for various strongly coupled supersymmetric field theories in arbitrary dimension with specific holographic duals. Furthermore, we have connected the closely related quantity $D_{3/2}$ which appears in the constitutive relation of the field theory's supersymmetry current to a universal fermionic absorption cross section result. In essence, this relation is very similar to the connection between shear viscosity and the universal scalar absorption cross section result in the dual space-time. In the latter case the relation has a direct field theoretical interpretation since the absorption cross section is given by the area of the horizon, which determines the entropy density of the field theory. In our case, the fermionic absorption cross section is also related to the area of a horizon, however, not in the original but in a conformally related spatially flat space-time. Therefore, the universal interpretation is not quite as 
straightforward as for $\eta / s$, however still very striking.

Clearly, the central questions which follow from our results relate to the universal result for the transport coefficient $D_{3/2}$, namely its range of applicability and possibly its significance towards real world systems, perhaps in the context of condensed matter or phenomenology, taking into account the interpretation of the setup~\cite{Kovtun:2003vj,Hoyos:2012dh}. 

It has been shown that the universality for $\eta / s$ holds for a very wide range of possible theories with Einstein gravity dual (note also the limitations of the universality in Gauss-Bonnet gravity~\cite{Buchel:2008vz} or for anisotropic setups~\cite{Erdmenger:2010xm,Erdmenger:2011tj,Rebhan:2011vd}). In comparison, for $D_{3/2}$, from the very beginning, we restricted to \emph{supersymmetric} field theories with (super-)gravity dual.

Furthermore, so far we ignored the possibility (or even necessity) of the coupling of our bulk gravitino to other fields like gauge fields, which could induce chemical potentials $\mu$ in the boundary theory. Generically, $D_{3/2}$ (or the supersound diffusion constant $D_s$) will depend on this $D_s = D_s \left(\frac{\mu}{T}\right)$~\cite{Gauntlett:2011mf,Gauntlett:2011wm,Kontoudi:2012mu}. Our calculations all refer to the transport coefficients' values in the limit $\mu \ll T$. Of course we would like to move away from the strict limit $\mu=0$ and study if for non-vanishing $\mu$ the shown universality still holds. So far we cannot tell if this is possible, but it is conceivable that the methods used (especially the transverse gravitino calculation in section~\ref{transverse}) are applicable for (at least) a perturbative treatment in $\mu/T$.

For checking other limitations of the universality one could, as for $\eta / s$, break the rotational symmetry of the boundary field theory as in~\cite{Erdmenger:2010xm} which would probably induce a temperature dependent deviation from the universal relation.

Another point that deserves attention is that in general fermions are not minimally coupled, but their equations of motion often involve Pauli terms which we have so far ignored. Without those, we may expect some universality to remain present. However, this is not the most generic situation. The fermionic universal absorption cross section result~\cite{Das:1996we} is given for minimally coupled fermions in spherically symmetric backgrounds (in arbitrary dimension) and we are not aware of direct extensions to non-minimally coupled fermions or minimally coupled fermions in rotating backgrounds. Such an extension would be important for studying e.g.\ the numerical results for $\mu\neq0$ in~\cite{Kontoudi:2012mu} along the lines of section~\ref{cross_sections} using consistent truncation arguments after embedding the setup into higher dimensional rotating backgrounds~\cite{Cvetic:1999xp}. Absorption cross sections of fermions in charged black hole backgrounds in four and five dimensions which do couple via 
Pauli terms have been studied in~\cite{Gubser:1997cm,Hosomichi:1997if}. On more general grounds, it might also be worth to study these cases from a purely gravitational point of view and see if a universal result in the form of~\cite{Das:1996we} may be extracted.

In using~\cite{Das:1996we}, we have been somewhat generous in using the distinction between black holes and black branes. We implicitly assumed and to some degree checked that the fermionic results of~\cite{Das:1996we} do hold also for the given branes. This is of course physically well motivated, but it is worth to examine this issue closer, as done for the scalar case in~\cite{Emparan:1997iv}. Actually, within our computations the contribution from additional extended brane dimensions seems to serve as the UV cutoff which we manually inserted at several places in our calculations. However~\cite{Emparan:1997iv} is more general in showing which space-times have low energy s-wave scalar absorption cross sections $\sigma_\text{abs,0}(0)=A$ and how this may be modified for others.

For non-conformal backgrounds, it would be interesting to also have some independent direct confirmation of the result~\eqref{Dp_brane_diffusion_constant} from an explicit computation along the lines of sections~\ref{transverse} or~\ref{longitudinal}. A Kubo formula for $D_{1/2}$ might be derived, so that we would be able to not only compute $D_s$ but also $D_\sigma$ for the non-conformal backgrounds.

It also seems possible to relate the present work to a supersymmetric extension of the fluid / gravity correspondence~\cite{Bhattacharyya:2008jc}. Initial progress in this direction for the BTZ black hole has been obtained in~\cite{Gentile:2013nha}.

Furthermore, the general appearance of the phonino excitation as the Goldstone fermion of supersymmetry breaking by temperature, which was first observed within holography in~\cite{Policastro:2008cx}, was given some more evidence in other dimensions. As done for the Wess-Zumino model~\cite{Kratzert:2003cr} and SQED~\cite{Kratzert:2002gh} it could be interesting to study this effect in other models, maybe even in simple supergravity models. For possible phenomenological implications we refer to~\cite{Hoyos:2012dh}.

\acknowledgments

We would like to thank M.~Ammon for initial collaboration on this project and for many further helpful comments. Furthermore, we thank W.~Buchm\"{u}ller, B.~Gout\'{e}raux, S.~Hartnoll, S.~Minwalla, Y.~Oz, K.~Skenderis and A.~Yarom, and in particular S.~Das and G.~Policastro for interesting discussions and correspondence on this and related subjects.

\bibliographystyle{JHEP}
\bibliography{Supersound_diffusion}

\end{document}